\documentclass[aps,showpacs,prb]{revtex4}
\usepackage{epsfig}
\begin{document}

\title{Prepyramid-to-pyramid transition of SiGe islands on Si(001)}

\author{A. Rastelli}
\email[]{A.Rastelli@fkf.mpg.de}
\affiliation{INFM and Dipartimento di Fisica ``A. Volta'',
Universit\`a degli Studi di Pavia, Via Bassi 6, I-27100 Pavia, Italy}
\altaffiliation{Present address: Max-Planck-Institut f\"ur Festk\"orperforschung, Heisenberstr. 1, D-70569 Stuttgart, Germany}
\author{H. Von K\"anel}
\affiliation{ETH Z\"urich, Laboratorium f\"ur Festk\"orperphysik,
CH-8093 Z\"urich, Switzerland.}
\affiliation{INFM and L-NESS Dipartimento di Fisica del Politecnico di Milano
Polo Regionale di Como Via Anzani 52 I-22100 Como, Italy}
\author{B. J. Spencer}
\affiliation{Department of Mathematics, State University of New York at Buffalo, Buffalo, New York 14260-2900, USA}
\author{J. Tersoff}
\affiliation{IBM Research Division, T. J. Watson Research Center, Yorktown Heights, New York 10598, USA}
\date{\today}

\begin{abstract}
The morphology of the first three-dimensional islands appearing during strained growth of SiGe alloys on Si(001) was investigated by scanning tunneling microscopy. High resolution images of individual islands and a statistical analysis of island shapes were used to reconstruct the evolution of the island shape as a function of size. As they grow, islands undergo a transition from completely unfacetted rough mounds (prepyramids) to partially \{105\} facetted islands and then they gradually evolve to \{105\} facetted pyramids. The results are in good agreement with the predictions of a recently proposed theoretical model.
\end{abstract}

\pacs{68.35.Bs; 68.37.Ef; 68.55.-a}

\maketitle

The first stages of spontaneous island formation and evolution during growth of SiGe on Si(001) have recently attracted much interest.~\cite{spencer97,tromp00,sutter00,vailionis00,rastelli02-1,chen97,floro99,berbezier02,tersoff02,shenoy02}
At typical growth temperatures, the Si$_{1-x}$Ge$_x$/Si(001) system displays a sequence of morphologies qualitatively independent of $x$, at least for $x\gtrsim0.2$.~\cite{floro98} In the case of low misfit strain ($x\lesssim 0.6$), individual islands have been observed to form without nucleation~\cite{tromp00,sutter00} from a quasi-periodic array of ripples as a result of a strain driven instability.~\cite{asaro72,grinfeld86,srolovitz89,spencer91}
Islands appear as unfacetted mounds  (``prepyramids"~\cite{vailionis00}) on a stressed wetting layer (WL) and then they evolve into facetted pyramids bounded by \{105\} facets as their size increases.

Recently we have shown that the observations reported in Refs.~\onlinecite{tromp00,sutter00,vailionis00,rastelli02-1} can be interpreted by assuming that the surface energy anisotropy of a SiGe layer on Si(001) has the following behavior:
(i) orientations in a neighborhood of (001) are permitted
so that (001) is not a true facet, (ii) \{105\} is a
facet orientation, and (iii) there is a range of unstable
orientations separating (001) and \{105\}.~\cite{tersoff02}
With these hypotheses, the equilibrium shape of an
island (in two dimensions) was calculated as a function
of size.
To simplify the calculations, \{105\} was not treated
as a true facet in our model, although it is one in the
real system.  The important point, which is captured
in the simplified model, is that there are a range of
unstable orientations separating (001) and \{105\}.

  Depending on the island volume $V$ there are shapes
corresponding to prepyramids and/or truncated pyramids with a rounded top (``T-pyramids").  The model predicts that there are only prepyramid islands for $V < V_1$, prepyramids and/or T-pyramids for $V_1 < V < V_3$, and only T-pyramids for $V > V_3$.

Most morphological studies of the first islands
appearing during SiGe growth on Si(001) have employed
techniques with low spatial resolution, such as
atomic force microscopy (AFM)~\cite{chen97,floro99,berbezier02}
or low energy electron microscopy (LEEM).~\cite{tromp00,sutter00}
The two exceptions studied pure Ge,~\cite{vailionis00}
where prepyramid islands are too small to resolve much detail,
or relatively dilute (25\%) Ge~\cite{sutter03} at relatively
low temperature (for this composition), where the
behavior may be rather different.~\cite{tersoff02}

We present here a scanning tunneling microscopy (STM) study of the pyramid precursors observed during growth of Si$_{1-x}$Ge$_x$ alloys on Si(001) in a composition range between $x=0.3$ and 0.5. A detailed statistical analysis of the shape as a function of island size is used to verify the predictions of the above model.
While a direct comparison of the data from three-dimensional (3D)
islands to the predictions for two-dimensional islands is not
possible, a qualitative comparison shows the striking similarities.

The samples for this study were grown by ultrahigh vacuum (UHV) magnetron sputter epitaxy and investigated by room temperature (RT) UHV-STM. Heavily doped Si(001) substrates with a nominal miscut $\lesssim 0.05^\circ$ were flash cleaned by alternating current heating in order to remove the native oxide. A Si buffer 100~nm thick was grown to achieve a clean surface. SiGe layers with different composition were obtained by tuning the relative power applied to two sputter guns operating simultaneously. SiGe was grown at a substrate temperature T$_s=600^\circ$C and a rate of about 0.08~nm/s. In some cases a short period of annealing at this temperature preceded the cooling of the sample to RT (at a rate of about $2^\circ$C/s) for STM characterization.

\begin{figure}
\epsfig{file=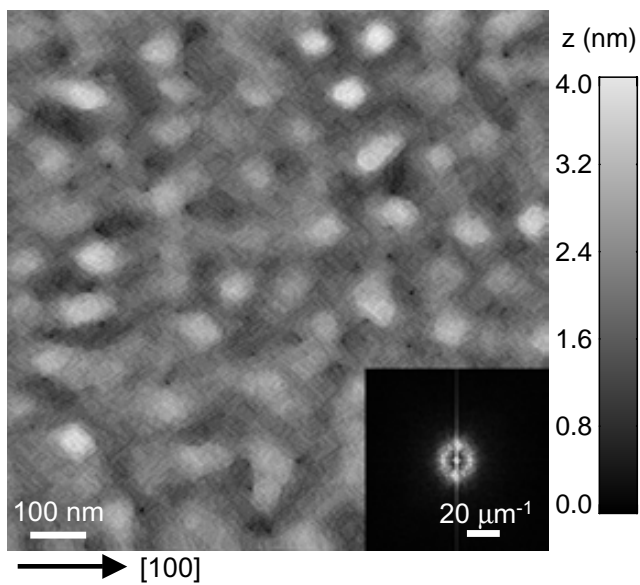, width=8.5cm}
\caption{STM image of a Si$_{0.7}$Ge$_{0.3}$ layer, 3.8~nm thick, grown on Si(001) at a substrate temperature of 600$^\circ$C and annealed for 20~s. Inset: average Fourier transform of six similar images. The observed ring corresponds to an average distance of 125~nm between mounds.}
\label{fig1}
\end{figure}

Figure~\ref{fig1} shows an STM image of a Si$_{0.7}$Ge$_{0.3}$ layer 3.8~nm thick annealed for 20~s at T$_s$. In spite of the very different growth technique employed here, our results are compatible with those obtained by Sutter and Lagally~\cite{sutter00} and Tromp {\it et al.}~\cite{tromp00} who used chemical vapor deposition (CVD) and studied the film morphology by means of LEEM. In fact, Fourier transforms of STM images of this sample show that islands start to develop from a ripple structure composed of cells with an average spacing of 125~nm (see inset of Fig.~\ref{fig1}). This finding is in good agreement with the value of 150~nm found in Ref.~\onlinecite{sutter00} for $x=0.3$, within the experimental uncertainties of  alloy composition.

\begin{figure}
\epsfig{file=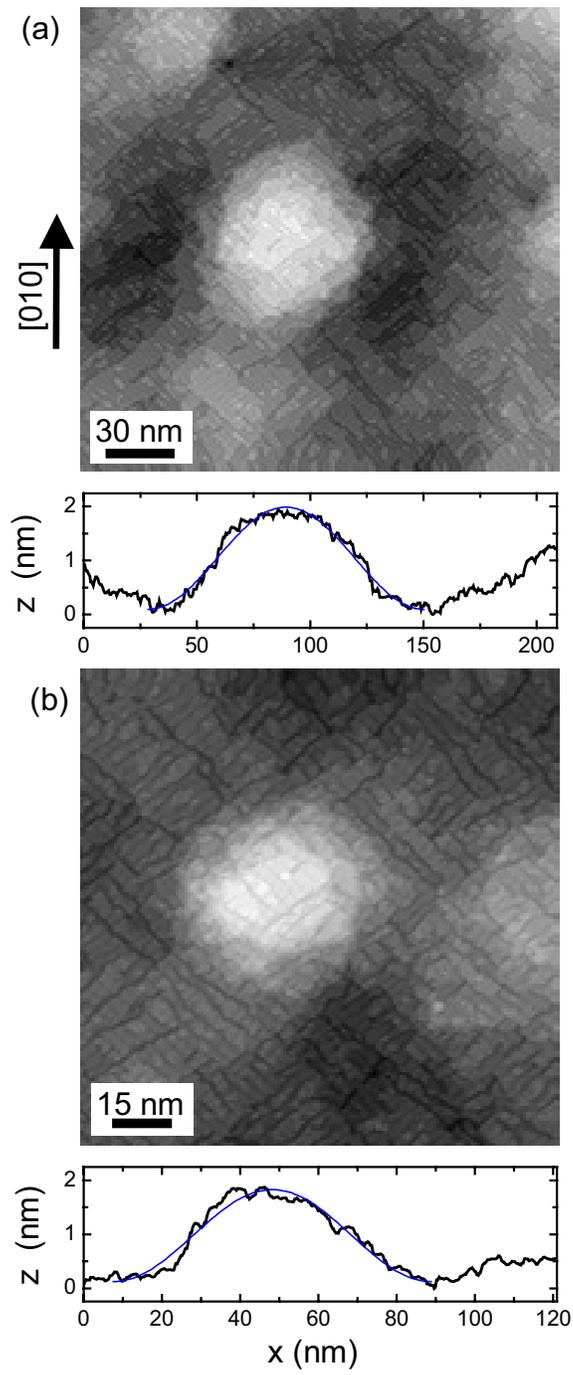, width=8.5cm}
\caption{STM images of Si$_{0.7}$Ge$_{0.3}$ (a) and Si$_{0.5}$Ge$_{0.5}$ (b) prepyramids and corresponding line scans through the center of the islands. Cosine fits of the line scans are also shown.}
\label{fig2}
\end{figure}

At higher resolution [Fig.~\ref{fig2}(a)] the mounds appear rounded and not regularly stepped, suggesting that, in contrast to clean Si(001), the (001) plane is not a facet for SiGe on Si(001). The contact angle between these prepyramids and the WL is zero, so it is difficult to identify the edge of the island base.

When the Ge fraction $x$ is increased to 0.5, the ripple structure is less clear, but still observable.
As in the case of lower misfit, islands appear as unfacetted mounds [Fig.~\ref{fig2}(b)]. As the composition is varied, prepyramids have qualitatively the same shape, but they occur at sizes decreasing with increasing $x$.
Small prepyramids were in fact observed also for nominally pure Ge grown on Si(001) by Vailionis {\it et al.}~\cite{vailionis00}

According to the model proposed in Refs.~\onlinecite{spencer97} and~\onlinecite{tersoff02}, the cross section of small prepyramids is well fitted by a cosine function. The two islands shown in Fig.~\ref{fig2} are close to the transition to the T-pyramid shape since small \{105\} facets are observable on their surface at higher resolution.~\cite{note105} For this reason the line scans deviate somewhat from the cosine fits. We observe that the WL and the surface of the islands show the same patched ($M\times N$) surface reconstruction with $M$ and $N$ increasing with decreasing $x$, as discussed in Refs.~\onlinecite{butz92,jernigan02} for the case of lower misfit alloys.

In order to obtain insight into the shape evolution, we examined samples featuring islands at different stages of evolution (prepyramids, T-pyramids and mature pyramids~\cite{notepyr}) and measured their shape as a function of size. Samples of this kind were obtained by suitably choosing the amount $\theta$ of deposited SiGe, having fixed the growth temperature to T$_s=600^\circ$C and composition to $x$=0.5.

\begin{figure}
\epsfig{file=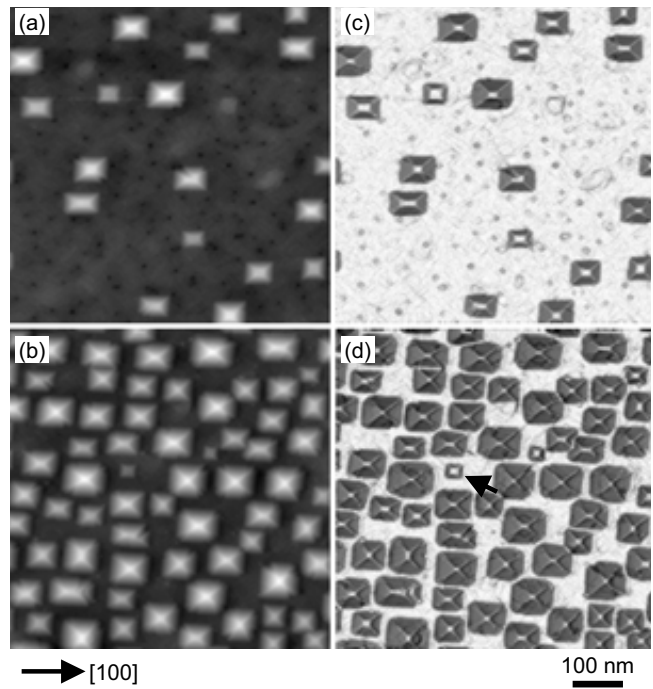, width=8.5cm}
\caption{STM images of Si$_{0.5}$Ge$_{0.5}$/Si(001) layers 1.5~nm thick, sample $S_{ld}$ (a) and 2.8~nm thick, sample $S_{hd}$ (b), annealed for 20~s at 600$^\circ$C. Vertical scale: 8~nm (a) and 10.5~nm (b). The right frames (c) and (d) show the same images, but with gray scale according to local surface slope: dark and light areas correspond to \{105\} facets and orientations close to (001), respectively. The arrow in (d) points at a small T-pyramid in the sample $S_{hd}$.}
\label{fig3}
\end{figure}

At $\theta=1.5$~nm the only 3D islands present on the surface are small prepyramids.
At $\theta=1.8$~nm a few facetted islands are observable together with partially facetted and unfacetted ones. However, a similar island distribution can be obtained after annealing a sample  with $\theta=1.5$~nm for 20~s at T$_s$ [see Fig.~\ref{fig3}(a) and (c)]. This observation is compatible with that of Jesson {\it et al.}~\cite{chen97,jesson96} and indicates that the critical thickness for the onset of 3D growth is kinetically determined, and that in general the whole film is  not in equilibrium.
The size distributions of the islands present in the above samples and in a sample with $\theta=1.9$~nm were found to be compatible. This whole set of samples will be referred to as $S_{ld}$, to denote that their surface is characterized by a relatively low density of islands (about $10^9$~cm$^{-2}$).
At later stages of growth the island density increases. Figure~\ref{fig3}(b) shows the surface of a sample ($S_{hd}$) obtained by growing $2.8$~nm of Si$_{0.5}$Ge$_{0.5}$ at $600^\circ$C and annealing it at this temperature for 20~s. Now, there are about $10^{10}$ islands per cm$^{2}$.
Besides mature pyramids (including islands close to the dome transition), small T-pyramids are observable also in this sample, as seen in Fig.~\ref{fig3}(d).

\begin{figure}
\epsfig{file=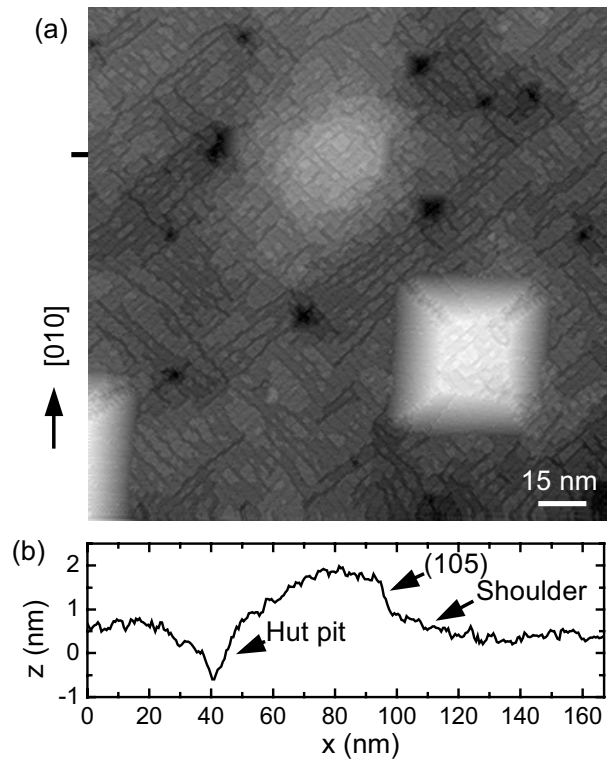, width=8.5cm}
\caption{(a) Magnification of two islands observed in Fig.~\ref{fig3}(a). Vertical scale: 4.5~nm. (b) Line scan of (a) along the [100] direction in correspondance with the black segment on the left side of the image.}
\label{fig4}
\end{figure}

Figure~\ref{fig4}(a) shows two islands at different stages of evolution and gives an indication on how a prepyramid transforms into a T-pyramid. The upper island displays a small (105) facet appearing on regions of its surface of the with the greatest slope, as revealed by the cross-sectional line scan shown in Fig.~\ref{fig4}(b). Both the region where the facet meets the rough top of the island and that where it meets the island base appear curved, as predicted by the model discussed in Ref.~\onlinecite{tersoff02}. The facet appears to sit on a shallow shoulder,~\cite{rastelli03} which shrinks for larger islands. This feature is in fact practically unobservable for the T-pyramid at the right lower corner of Fig.~\ref{fig4}(a), consistent with the model.

In order to describe the shape evolution as a function of island size,
for each island we determined several parameters by means of a dedicated software program.
As noted above, the bases of prepyramids are not sharply delineated. Roughness of the surrounding WL and ``hut pits"~\cite{goldfarb97,jesson00} [visible as small dark spots in Fig.~\ref{fig3}(c), and in more detail in Fig.~\ref{fig4}] make it difficult to identify the island edge. For this reason, the island base perimeters were defined by means of a contour plot at a reference height $h_0$.
$h_0$ was chosen 0.5~nm above the average height at which \{105\} facets belonging to pyramids meet the WL. In this way the shallow shoulder (see Fig.~\ref{fig4}) at the base of T-pyramid was not included in the analysis. Island heights were measured with respect to $h_0$ and volumes were calculated by integration.
Two further parameters were measured for each island: (i) the area $A_{(105)}$ of the island surface occupied by \{105\} facets (as described in Ref.~\onlinecite{rastelli02-2}), (ii) the area $A_{(001)}$ of the regions of the island surface with inclination less than 6 degrees. For prepyramids and T-pyramids, the latter parameter gives a measure of the total area and of the area of the rounded top, respectively.
Islands that were too asymmetric, coalescing or poorly defined were excluded from the analysis in order to reduce the measurement uncertainty.


We note that pits in the WL locally modify the strain configuration, possibly leading to the occurrence of asymmetric islands.  At present we are not able to exclude that these pits are correlated to impurities~\cite{deng98} introduced at the SiGe/Si interface during the necessary growth interruption before initiation of SiGe growth. If this is the case, it may be possible to produce samples with a flatter WL, allowing the uncertainties on the size measurement of prepyramids to be reduced and the model assumptions to be better approached. Finally the cooling of the sample may affect the shape of the islands and change it from its ideal equilibrium configuration.~\cite{ross99}

\begin{figure}
\epsfig{file=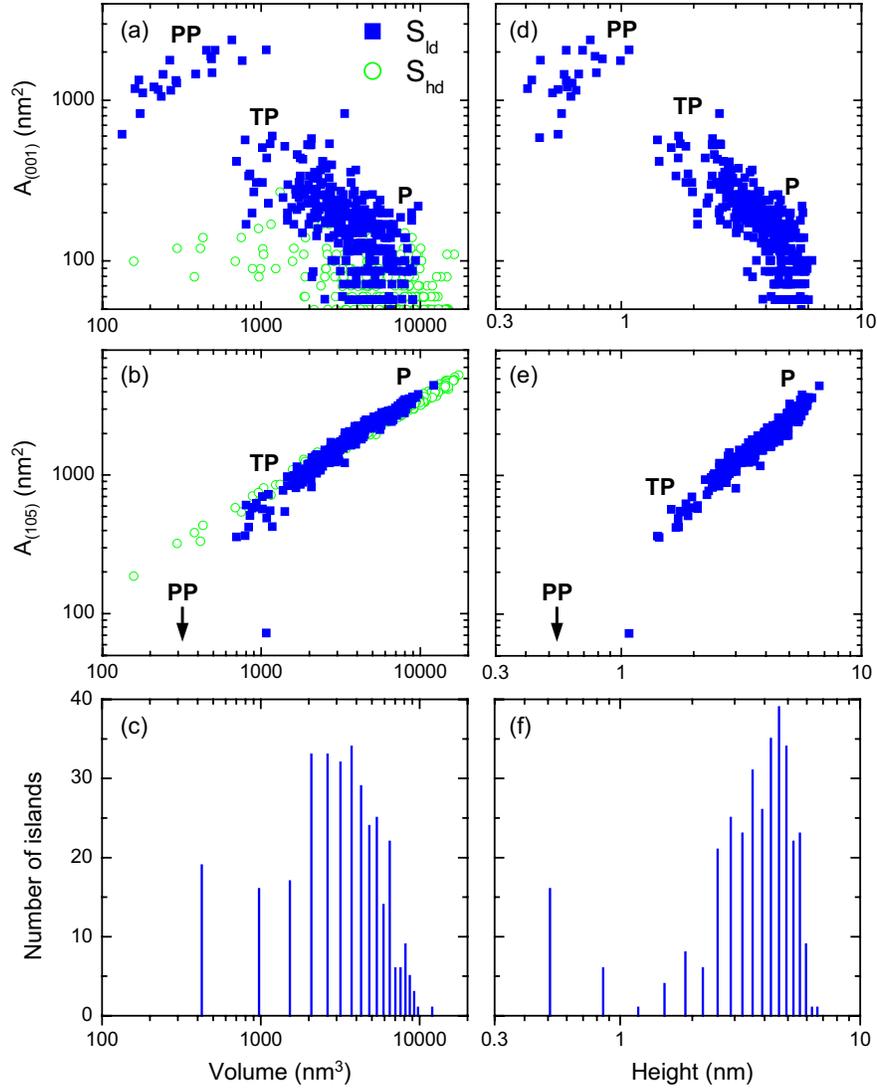, width=13cm}
\caption{Scatter plots for 331 islands (solid squares) on a set of four samples ($S_{ld}$) obtained by depositing Si$_{0.5}$Ge$_{0.5}$ layers with thickness in the range 1.5-1.9~nm on a Si(001) substrate at 600$^\circ$C. The area of the surface occupied by \{105\} facets ($A_{(105)}$) and that of regions with orientation close to (001), ($A_{(001)}$) are plotted vs island volume (a), (b) and height (d), (e). Plots of $A_{(001)}$ and $A_{(105)}$  vs volume for 396 islands contained in sample $S_{hd}$ (see Fig.~\ref{fig3} for details) are also shown in (a) and (b) as open circles. Volume and height distributions for samples $S_{ld}$ are shown in (c) and (f), respectively.
With increasing size, islands transform from unfacetted prepyramids (PP) to partially facetted T-pyramids (TP) and then gradually to mature pyramids (P).}
\label{fig5}
\end{figure}

Figure~\ref{fig5} illustrates the transition of prepyramids to T-pyramids and their gradual evolution to mature pyramids~\cite{notepyr} with increasing size.
Prepyramids are characterized by a large $A_{(001)}$ and a negligible $A_{(105)}$, while the surface of T-pyramids contains both, \{105\} facets and areas with orientation close to (001).

Focusing initially on the samples $S_{ld}$,
Figs.~\ref{fig5}(a) and (b) show that the smallest 3D islands are prepyramids and that these islands ``survive" in a limited range of volumes. The transition occurs at a volume of  about 1000~nm$^3$, where \{105\} facets appear [Fig.~\ref{fig5}(b)]. Figure~\ref{fig5}(a) suggests that a discontinuous drop in $A_{(001)}$ occurs at the transition.
As the island volume increases beyond the critical size, we observe that: (i) the  rounded top of T-pyramids smoothly shrinks, so that the larger the T-pyramid, the sharper is its apex [Fig.~\ref{fig5}(a)] and (ii) the area of \{105\} facets increases monotonically as nearly a power law [Fig.~\ref{fig5}(b)].

Qualitatively, the same considerations apply to the plots of $A_{(001)}$ and $A_{(105)}$ vs height [Figs.~\ref{fig5}(d) and (e)]. However, while the range of volumes of prepyramids overlaps that of T-pyramids [Fig.~\ref{fig5}(a)], there is a clear separation between the range of heights of prepyramids and of T-pyramids [Fig.~\ref{fig5}(d)].

Thus from the data the following scenario is suggested.
At small volume an island is a prepyramid.
At some critical volume the prepyramid undergoes an abrupt transition to
a T-pyramid, accompanied by a discontinuous jump in the area of \{105\}
regions and a discontinous jump in the height of the island. Upon
further growth, the island reduces the amount of (001) area while
increasing the amount of \{105\} regions.

Turning now to the data from sample $S_{hd}$,
Figs.~\ref{fig5}(a) and (b) show that
these islands
have size distributions substantially different from
those of samples $S_{ld}$.
 The smallest islands observed in sample $S_{hd}$ [such as that pointed at by the arrow in Fig.~\ref{fig3}(d)] are T-pyramids with a relatively large flat top and with a volume appreciably smaller than 1000~nm$^3$.
No unfacetted islands were seen in sample $S_{hd}$,
down to the smallest size that we can reliably identify.
The model presented in Ref.~\onlinecite{tersoff02} offers an interesting interpretation of this observation, as discussed below.

\begin{figure}
\epsfig{file=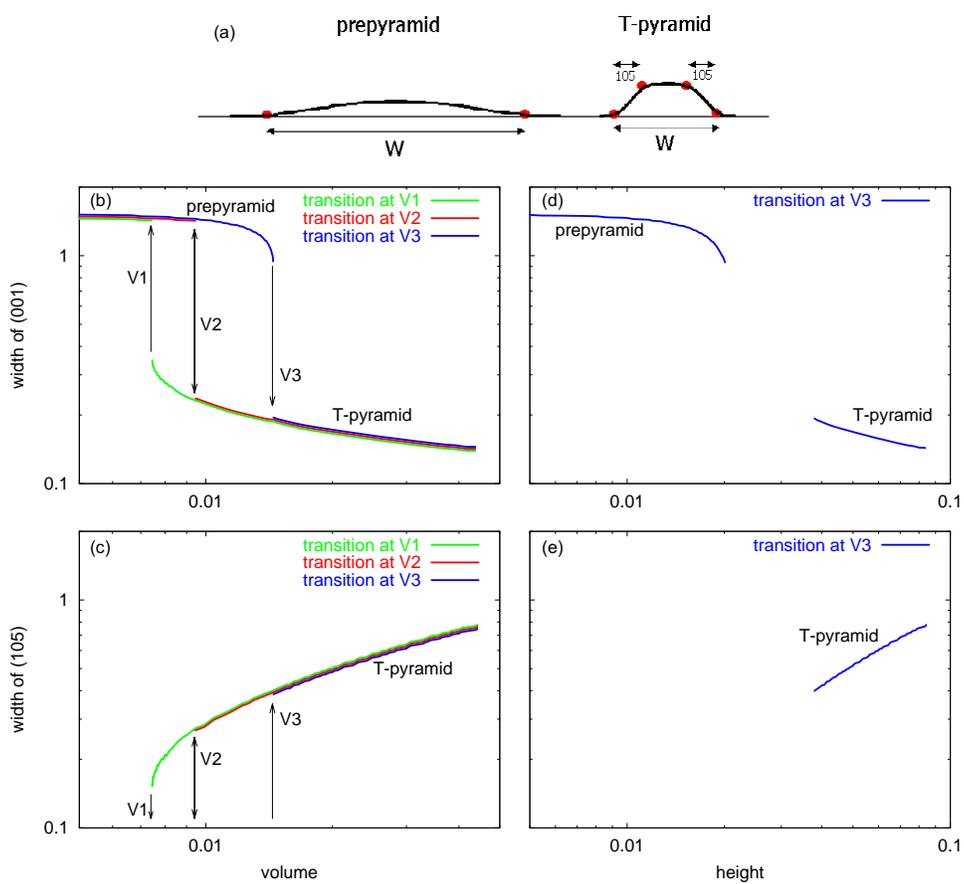, width=13cm}
\caption{Schematic representation of the predicted equilibrium shapes (a) and width of (001) and \{105\} regions vs island volume (b), (c) and height (d), (e) from theoretical model.}
\label{fig6}
\end{figure}

The theoretical island shapes from Ref.~\onlinecite{tersoff02} were analyzed
to determine
shape parameters analogous to those measured in Fig.~\ref{fig5}.
We determined the ``apparent'' width of prepyramid and
T-pyramid theory shapes from the points of maximum concave
curvature [see schematic in Fig.~\ref{fig6}(a)].  For
T-pyramids, the widths of the (001) and \{105\} regions
were determined from the point of maximum convex
curvature.

Figures~\ref{fig6}(b) and (c) show the width of the (001) and \{105\} regions as a function of the nondimensional island volume.  Prepyramids extending from $V=0$ to $V=V_3$ have no \{105\} regions.  T-pyramid solutions occur for $V>V_1$ with the amount of \{105\} increasing
monotonically with volume. As the island grows, the transition from prepyramid to T-pyramid is energetically favorable for $V>V_2$, and has no barrier for $V>V_3$.
The inverse transition, for a shrinking island to change from T-pyramid to prepyramid, is favorable for $V<V_2$ and has no barrier for $V<V_1$.
Throughout the range $V_2<V<V_3$ the transition
from prepyramid to T-pyramid is characterized by
a jump in the \{105\} width. This
behavior is qualitatively similar to the $S_{ld}$ data in
Fig.~\ref{fig5}(b) which shows the abrupt transition
in the \{105\} area when the island volume
is near 1000 nm$^3$.

If the activation energies are large compared to thermal fluctuations,
the transition from prepyramid to T-pyramid will occur at $V_3$, while
the transition from
T-pyramid to prepyramid will occur at $V_1$.  Consequently,
the prepyramid/T-pyramid transition will demonstrate
hysteresis behavior as illustrated in Figs.~\ref{fig6}(b) and (c).
Thus, the model suggests that the observed difference in the transition size between samples $S_{ld}$ and $S_{hd}$ [see Figs.~\ref{fig5}(a) and (b)] can be  understood if islands in samples $S_{ld}$ are growing, while the small T-pyramids in sample $S_{hd}$ [see Fig.~\ref{fig3}(d)] are in the process of shrinking.
This is a probable scenario, since in sample $S_{hd}$ large pyramids act as sink of material and grow at the expense of smaller islands. This is analogous to the anomalous ripening process observed by Ross {\it et al.}~\cite{ross98-1} in the case of the transition from pyramid to dome.
Then Figs.~\ref{fig5}(a) and (b) allow a direct comparison of the volume distribution in the case of growing and shrinking T-pyramids in qualitative agreement with the model. The $S_{ld}$ data would correspond to the ``forward" transition in which T-pyramids occur for $V>V_3$, while the $S_{hd}$ data sample some of the ``backward" transition in which T-pyramids can also occur for $V_1< V <V_3$.~\cite{note_S_hd}

Figures~\ref{fig6}(d) and (e) show the
plots of the ``area'' (width in this two dimensional calculation) as a function of height
for the theoretical solutions. Here we plot the
shape parameters from the theory, taking the transition
to occur at $V_3$.   As with the data,
the width of the prepyramid solutions is relatively
constant for small island heights.  The transition
to T-pyramids is marked by a discontinuous jump
in the height, leading to well-separated curves
for prepyramids and T-pyramids.  Finally, as
T-pyramids grow, the area of the \{105\} regions
grow while the area of the (001) region shrinks.

In conclusion, we have investigated by STM the early stages of island formation and faceting during sputter deposition of SiGe alloys on Si(001). Besides confirming previous observations obtained with different growth and characterization techniques, our study reveals new details of the shape of the pyramid precursors in agreement with the predictions of a model recently proposed.~\cite{tersoff02} A statistical analysis of the island shape as a function of size allows us to follow the morphological transition in which unfacetted prepyramids transform into  facetted pyramids and {\it vice versa}. Our model is able to grasp the main observations reported here.

A.R. acknowledges for support and hospitality the ETH of Z\"urich (Switzerland) and the ORC of the Technical University of Tampere (Finland). Fruitful discussions with Chiu Cheng-Hsin and D. Chrastina are also acknowledged. This work is supported by the National Science Foundation under Grant No. NIRT  DMR-0102794 (B.J.S.).

\bibliography{pp}

\begin{thebibliography}{29}
\expandafter\ifx\csname natexlab\endcsname\relax\def\natexlab#1{#1}\fi
\expandafter\ifx\csname bibnamefont\endcsname\relax
  \def\bibnamefont#1{#1}\fi
\expandafter\ifx\csname bibfnamefont\endcsname\relax
  \def\bibfnamefont#1{#1}\fi
\expandafter\ifx\csname citenamefont\endcsname\relax
  \def\citenamefont#1{#1}\fi
\expandafter\ifx\csname url\endcsname\relax
  \def\url#1{\texttt{#1}}\fi
\expandafter\ifx\csname urlprefix\endcsname\relax\def\urlprefix{URL }\fi
\providecommand{\bibinfo}[2]{#2}
\providecommand{\eprint}[2][]{\url{#2}}

\bibitem[{\citenamefont{Spencer and Tersoff}(1997)}]{spencer97}
\bibinfo{author}{\bibfnamefont{B.~J.} \bibnamefont{Spencer}} \bibnamefont{and}
  \bibinfo{author}{\bibfnamefont{J.}~\bibnamefont{Tersoff}},
  \bibinfo{journal}{Phys.\ Rev.\ Lett.} \textbf{\bibinfo{volume}{79}},
  \bibinfo{pages}{4858} (\bibinfo{year}{1997}).

\bibitem[{\citenamefont{Tromp et~al.}(2000)\citenamefont{Tromp, Ross, and
  Reuter}}]{tromp00}
\bibinfo{author}{\bibfnamefont{R.~M.} \bibnamefont{Tromp}},
  \bibinfo{author}{\bibfnamefont{F.~M.} \bibnamefont{Ross}}, \bibnamefont{and}
  \bibinfo{author}{\bibfnamefont{M.~C.} \bibnamefont{Reuter}},
  \bibinfo{journal}{Phys.\ Rev.\ Lett.} \textbf{\bibinfo{volume}{84}},
  \bibinfo{pages}{4641} (\bibinfo{year}{2000}).

\bibitem[{\citenamefont{Sutter and Lagally}(2000)}]{sutter00}
\bibinfo{author}{\bibfnamefont{P.}~\bibnamefont{Sutter}} \bibnamefont{and}
  \bibinfo{author}{\bibfnamefont{M.~G.} \bibnamefont{Lagally}},
  \bibinfo{journal}{Phys.\ Rev.\ Lett.} \textbf{\bibinfo{volume}{84}},
  \bibinfo{pages}{4637} (\bibinfo{year}{2000}).

\bibitem[{\citenamefont{Vailionis et~al.}(2000)\citenamefont{Vailionis, Cho,
  Glass, Desjardins, Cahill, and Greene}}]{vailionis00}
\bibinfo{author}{\bibfnamefont{A.}~\bibnamefont{Vailionis}},
  \bibinfo{author}{\bibfnamefont{B.}~\bibnamefont{Cho}},
  \bibinfo{author}{\bibfnamefont{G.}~\bibnamefont{Glass}},
  \bibinfo{author}{\bibfnamefont{P.}~\bibnamefont{Desjardins}},
  \bibinfo{author}{\bibfnamefont{D.~G.} \bibnamefont{Cahill}},
  \bibnamefont{and} \bibinfo{author}{\bibfnamefont{J.~E.}
  \bibnamefont{Greene}}, \bibinfo{journal}{Phys.\ Rev.\ Lett.}
  \textbf{\bibinfo{volume}{85}}, \bibinfo{pages}{3672} (\bibinfo{year}{2000}).

\bibitem[{\citenamefont{Rastelli et~al.}(2002)\citenamefont{Rastelli, Kummer,
  and von K\"anel}}]{rastelli02-1}
\bibinfo{author}{\bibfnamefont{A.}~\bibnamefont{Rastelli}},
  \bibinfo{author}{\bibfnamefont{M.}~\bibnamefont{Kummer}}, \bibnamefont{and}
  \bibinfo{author}{\bibfnamefont{H.}~\bibnamefont{von K\"anel}},
  \bibinfo{journal}{Mat.\ Res.\ Soc.\ Symp.\ Proc.}
  \textbf{\bibinfo{volume}{696}}, \bibinfo{pages}{N5.4.1.}
  (\bibinfo{year}{2002}).

\bibitem[{\citenamefont{Chen et~al.}(1997)\citenamefont{Chen, Jesson,
  Pennycook, Thundat, and Warmack}}]{chen97}
\bibinfo{author}{\bibfnamefont{K.~M.} \bibnamefont{Chen}},
  \bibinfo{author}{\bibfnamefont{D.~E.} \bibnamefont{Jesson}},
  \bibinfo{author}{\bibfnamefont{S.~J.} \bibnamefont{Pennycook}},
  \bibinfo{author}{\bibfnamefont{T.}~\bibnamefont{Thundat}}, \bibnamefont{and}
  \bibinfo{author}{\bibfnamefont{R.~J.} \bibnamefont{Warmack}},
  \bibinfo{journal}{Phys.\ Rev.\ B} \textbf{\bibinfo{volume}{56}},
  \bibinfo{pages}{R1700} (\bibinfo{year}{1997}).

\bibitem[{\citenamefont{Floro et~al.}(1999)\citenamefont{Floro, Chason, Freund,
  Twesten, Hwang, and Lucadamo}}]{floro99}
\bibinfo{author}{\bibfnamefont{J.~A.} \bibnamefont{Floro}},
  \bibinfo{author}{\bibfnamefont{E.}~\bibnamefont{Chason}},
  \bibinfo{author}{\bibfnamefont{L.~B.} \bibnamefont{Freund}},
  \bibinfo{author}{\bibfnamefont{R.~D.} \bibnamefont{Twesten}},
  \bibinfo{author}{\bibfnamefont{R.~Q.} \bibnamefont{Hwang}}, \bibnamefont{and}
  \bibinfo{author}{\bibfnamefont{G.~A.} \bibnamefont{Lucadamo}},
  \bibinfo{journal}{Phys.\ Rev.\ B} \textbf{\bibinfo{volume}{59}},
  \bibinfo{pages}{1990} (\bibinfo{year}{1999}).

\bibitem[{\citenamefont{{Berbezier} et~al.}(2002)\citenamefont{{Berbezier},
  {Ronda}, and {Portavoce}}}]{berbezier02}
\bibinfo{author}{\bibfnamefont{I.}~\bibnamefont{{Berbezier}}},
  \bibinfo{author}{\bibfnamefont{A.}~\bibnamefont{{Ronda}}}, \bibnamefont{and}
  \bibinfo{author}{\bibfnamefont{A.}~\bibnamefont{{Portavoce}}},
  \bibinfo{journal}{J. Phys.: Condens. Matter} \textbf{\bibinfo{volume}{14}},
  \bibinfo{pages}{8283} (\bibinfo{year}{2002}).

\bibitem[{\citenamefont{Tersoff et~al.}(2002)\citenamefont{Tersoff, Spencer,
  Rastelli, and von K\"anel}}]{tersoff02}
\bibinfo{author}{\bibfnamefont{J.}~\bibnamefont{Tersoff}},
  \bibinfo{author}{\bibfnamefont{B.~J.} \bibnamefont{Spencer}},
  \bibinfo{author}{\bibfnamefont{A.}~\bibnamefont{Rastelli}}, \bibnamefont{and}
  \bibinfo{author}{\bibfnamefont{H.}~\bibnamefont{von K\"anel}},
  \bibinfo{journal}{Phys.\ Rev.\ Lett.} \textbf{\bibinfo{volume}{89}},
  \bibinfo{pages}{196104} (\bibinfo{year}{2002}).

\bibitem[{\citenamefont{Shenoy et~al.}(2002)\citenamefont{Shenoy, Ciobanu, and
  Freund}}]{shenoy02}
\bibinfo{author}{\bibfnamefont{V.~B.} \bibnamefont{Shenoy}},
  \bibinfo{author}{\bibfnamefont{C.~V.} \bibnamefont{Ciobanu}},
  \bibnamefont{and} \bibinfo{author}{\bibfnamefont{L.~B.}
  \bibnamefont{Freund}}, \bibinfo{journal}{Appl. Phys. Lett.}
  \textbf{\bibinfo{volume}{81}}, \bibinfo{pages}{364} (\bibinfo{year}{2002}).

\bibitem[{\citenamefont{Floro et~al.}(1998)\citenamefont{Floro, Lucadamo,
  Chason, Freund, Sinclair, Twesten, and Hwang}}]{floro98}
\bibinfo{author}{\bibfnamefont{J.~A.} \bibnamefont{Floro}},
  \bibinfo{author}{\bibfnamefont{G.~A.} \bibnamefont{Lucadamo}},
  \bibinfo{author}{\bibfnamefont{E.}~\bibnamefont{Chason}},
  \bibinfo{author}{\bibfnamefont{L.~B.} \bibnamefont{Freund}},
  \bibinfo{author}{\bibfnamefont{M.}~\bibnamefont{Sinclair}},
  \bibinfo{author}{\bibfnamefont{R.~D.} \bibnamefont{Twesten}},
  \bibnamefont{and} \bibinfo{author}{\bibfnamefont{R.~Q.} \bibnamefont{Hwang}},
  \bibinfo{journal}{Phys.\ Rev.\ Lett.} \textbf{\bibinfo{volume}{80}},
  \bibinfo{pages}{4717} (\bibinfo{year}{1998}).

\bibitem[{\citenamefont{Asaro and Tiller}(1972)}]{asaro72}
\bibinfo{author}{\bibfnamefont{R.~J.} \bibnamefont{Asaro}} \bibnamefont{and}
  \bibinfo{author}{\bibfnamefont{W.~A.} \bibnamefont{Tiller}},
  \bibinfo{journal}{Met. Trans.} \textbf{\bibinfo{volume}{3}},
  \bibinfo{pages}{1789} (\bibinfo{year}{1972}).

\bibitem[{\citenamefont{Grinfeld}(1986)}]{grinfeld86}
\bibinfo{author}{\bibfnamefont{M.~A.} \bibnamefont{Grinfeld}},
  \bibinfo{journal}{Sov. Phys. Dokl.} \textbf{\bibinfo{volume}{31}},
  \bibinfo{pages}{831} (\bibinfo{year}{1986}).

\bibitem[{\citenamefont{Srolovitz}(1989)}]{srolovitz89}
\bibinfo{author}{\bibfnamefont{D.~J.} \bibnamefont{Srolovitz}},
  \bibinfo{journal}{Acta Metall.} \textbf{\bibinfo{volume}{37}},
  \bibinfo{pages}{621} (\bibinfo{year}{1989}).

\bibitem[{\citenamefont{{Spencer} et~al.}(1991)\citenamefont{{Spencer},
  {Voorhees}, and {Davis}}}]{spencer91}
\bibinfo{author}{\bibfnamefont{B.~J.} \bibnamefont{{Spencer}}},
  \bibinfo{author}{\bibfnamefont{P.~W.} \bibnamefont{{Voorhees}}},
  \bibnamefont{and} \bibinfo{author}{\bibfnamefont{S.~H.}
  \bibnamefont{{Davis}}}, \bibinfo{journal}{Phys.\ Rev.\ Lett.}
  \textbf{\bibinfo{volume}{67}}, \bibinfo{pages}{3696} (\bibinfo{year}{1991}).

\bibitem[{\citenamefont{Sutter et~al.}(2003)\citenamefont{Sutter, Zahl, and
  Sutter}}]{sutter03}
\bibinfo{author}{\bibfnamefont{P.}~\bibnamefont{Sutter}},
  \bibinfo{author}{\bibfnamefont{P.}~\bibnamefont{Zahl}}, \bibnamefont{and}
  \bibinfo{author}{\bibfnamefont{E.}~\bibnamefont{Sutter}},
  \bibinfo{journal}{Appl. Phys. Lett.} \textbf{\bibinfo{volume}{82}},
  \bibinfo{pages}{3454} (\bibinfo{year}{2003}).

\bibitem[{not({\natexlab{a}})}]{note105}
\bibinfo{note}{Even small \{105\} facets are easily recognizable in STM images
  with a lateral resolution of at least 0.5~nm, due to their distinctive
  surface reconstruction.}

\bibitem[{\citenamefont{Butz and Kampers}(1992)}]{butz92}
\bibinfo{author}{\bibfnamefont{R.}~\bibnamefont{Butz}} \bibnamefont{and}
  \bibinfo{author}{\bibfnamefont{S.}~\bibnamefont{Kampers}},
  \bibinfo{journal}{Appl. Phys. Lett.} \textbf{\bibinfo{volume}{61}},
  \bibinfo{pages}{1307} (\bibinfo{year}{1992}).

\bibitem[{\citenamefont{Jernigan and Thompson}(2002)}]{jernigan02}
\bibinfo{author}{\bibfnamefont{G.~G.} \bibnamefont{Jernigan}} \bibnamefont{and}
  \bibinfo{author}{\bibfnamefont{P.}~\bibnamefont{Thompson}},
  \bibinfo{journal}{Surf. Sci.} \textbf{\bibinfo{volume}{516}},
  \bibinfo{pages}{207} (\bibinfo{year}{2002}).

\bibitem[{not({\natexlab{b}})}]{notepyr}
\bibinfo{note}{By ``mature'' pyramids we mean T-pyramids whose top has a width
  comparable to or smaller than the lateral resolution of our measurement.}

\bibitem[{\citenamefont{Jesson et~al.}(1996)\citenamefont{Jesson, Chen,
  Pennycook, Thundat, and Warmack}}]{jesson96}
\bibinfo{author}{\bibfnamefont{D.~E.} \bibnamefont{Jesson}},
  \bibinfo{author}{\bibfnamefont{K.~M.} \bibnamefont{Chen}},
  \bibinfo{author}{\bibfnamefont{S.~J.} \bibnamefont{Pennycook}},
  \bibinfo{author}{\bibfnamefont{T.}~\bibnamefont{Thundat}}, \bibnamefont{and}
  \bibinfo{author}{\bibfnamefont{R.~J.} \bibnamefont{Warmack}},
  \bibinfo{journal}{Phys.\ Rev.\ Lett.} \textbf{\bibinfo{volume}{77}},
  \bibinfo{pages}{1330} (\bibinfo{year}{1996}).

\bibitem[{\citenamefont{Rastelli and von K\"anel}(2003)}]{rastelli03}
\bibinfo{author}{\bibfnamefont{A.}~\bibnamefont{Rastelli}} \bibnamefont{and}
  \bibinfo{author}{\bibfnamefont{H.}~\bibnamefont{von K\"anel}},
  \bibinfo{journal}{Surf.\ Sci.} \textbf{\bibinfo{volume}{532-535}},
  \bibinfo{pages}{769} (\bibinfo{year}{2003}).

\bibitem[{\citenamefont{Goldfarb et~al.}(1997)\citenamefont{Goldfarb, Hayden,
  Owen, and Briggs}}]{goldfarb97}
\bibinfo{author}{\bibfnamefont{I.}~\bibnamefont{Goldfarb}},
  \bibinfo{author}{\bibfnamefont{P.~T.} \bibnamefont{Hayden}},
  \bibinfo{author}{\bibfnamefont{J.~H.~G.} \bibnamefont{Owen}},
  \bibnamefont{and} \bibinfo{author}{\bibfnamefont{G.~A.~D.}
  \bibnamefont{Briggs}}, \bibinfo{journal}{Phys.\ Rev.\ Lett.}
  \textbf{\bibinfo{volume}{78}}, \bibinfo{pages}{3959} (\bibinfo{year}{1997}).

\bibitem[{\citenamefont{Jesson et~al.}(2000)\citenamefont{Jesson, K\"astner,
  and Voigtl\"ander}}]{jesson00}
\bibinfo{author}{\bibfnamefont{D.~E.} \bibnamefont{Jesson}},
  \bibinfo{author}{\bibfnamefont{M.}~\bibnamefont{K\"astner}},
  \bibnamefont{and}
  \bibinfo{author}{\bibfnamefont{B.}~\bibnamefont{Voigtl\"ander}},
  \bibinfo{journal}{Phys.\ Rev.\ Lett.} \textbf{\bibinfo{volume}{84}},
  \bibinfo{pages}{330} (\bibinfo{year}{2000}).

\bibitem[{\citenamefont{Rastelli and von K\"anel}(2002)}]{rastelli02-2}
\bibinfo{author}{\bibfnamefont{A.}~\bibnamefont{Rastelli}} \bibnamefont{and}
  \bibinfo{author}{\bibfnamefont{H.}~\bibnamefont{von K\"anel}},
  \bibinfo{journal}{Surf.\ Sci.\ Lett.} \textbf{\bibinfo{volume}{515}},
  \bibinfo{pages}{L493} (\bibinfo{year}{2002}).

\bibitem[{\citenamefont{Deng and Krishnamurthy}(1998)}]{deng98}
\bibinfo{author}{\bibfnamefont{X.}~\bibnamefont{Deng}} \bibnamefont{and}
  \bibinfo{author}{\bibfnamefont{M.}~\bibnamefont{Krishnamurthy}},
  \bibinfo{journal}{Phys.\ Rev.\ Lett.} \textbf{\bibinfo{volume}{81}},
  \bibinfo{pages}{1473} (\bibinfo{year}{1998}).

\bibitem[{\citenamefont{Ross et~al.}(1999)\citenamefont{Ross, Tromp, and
  Reuter}}]{ross99}
\bibinfo{author}{\bibfnamefont{F.~M.} \bibnamefont{Ross}},
  \bibinfo{author}{\bibfnamefont{R.~M.} \bibnamefont{Tromp}}, \bibnamefont{and}
  \bibinfo{author}{\bibfnamefont{M.~C.} \bibnamefont{Reuter}},
  \bibinfo{journal}{Science} \textbf{\bibinfo{volume}{286}},
  \bibinfo{pages}{1931} (\bibinfo{year}{1999}).

\bibitem[{\citenamefont{Ross et~al.}(1998)\citenamefont{Ross, Tersoff, and
  Tromp}}]{ross98-1}
\bibinfo{author}{\bibfnamefont{F.~M.} \bibnamefont{Ross}},
  \bibinfo{author}{\bibfnamefont{J.}~\bibnamefont{Tersoff}}, \bibnamefont{and}
  \bibinfo{author}{\bibfnamefont{R.~M.} \bibnamefont{Tromp}},
  \bibinfo{journal}{Phys.\ Rev.\ Lett.} \textbf{\bibinfo{volume}{80}},
  \bibinfo{pages}{984} (\bibinfo{year}{1998}).

\bibitem[{not({\natexlab{c}})}]{note_S_hd}
\bibinfo{note}{An alternative explanation for the different island
  size-distributions in samples $S_{ld}$ and $S_{hd}$ is that, when the island
  density is large, the elastic interaction between islands can reduce the
  transition size, as observed by Floro {\it et al.}~\cite{floro98} for the
  transition from SiGe pyramids to domes. Following this argument, the smallest
  T-pyramids seen in sample $S_{hd}$ could be the result of the transition of
  prepyramids surrounded by larger T-pyramids. Further experiments may clarify
  the issue.}

\end{thebibliography}

\newpage

\end{document}